\documentclass[a4paper,12pt]{article}
\usepackage{amssymb}
\usepackage{epsfig}
\usepackage{float}
\usepackage{graphicx}
\usepackage{amsmath}
\newcommand{\nua}[1]{\ensuremath{\rlap
           {\kern-2.5pt\ensuremath
           {\overset{\scriptscriptstyle(-)}{\phantom{\nu}}}}
           {\ensuremath{{\nu}_{#1}}}}}

\begin{document}
\begin{center}
{\bf Neutrino oscillations:                                                                                           from an historical perspective to the present status }

\end{center}
\begin{center}
S.  Bilenky
\end{center}
\begin{center}
{\em  Joint Institute for Nuclear Research, Dubna, R-141980,
Russia\\}
{\em TRIUMF
4004 Wesbrook Mall,
Vancouver BC, V6T 2A3
Canada\\}
\end{center}
\begin{abstract}
The history of neutrino mixing and oscillations is briefly presented. Basics of neutrino mixing and oscillations and convenient formalism of neutrino oscillations in vacuum is given. The role of neutrino in the Standard Model and the Weinberg mechanism of the generation of the Majorana neutrino masses are discussed.
\end{abstract}

\section{Introduction. On the history of neutrino oscillations}

Discovery of the neutrino oscillations in the atmospheric Super-Kamiokande \cite{Fukuda:1998ah}, solar SNO \cite{Ahmad:2002jz} and reactor KamLAND \cite{Eguchi:2002dm} experiments was a first  evidence in favor of a beyond the Standard Model physics in particle physics. Neutrino oscillations were further studied in the long baseline accelerator  K2K \cite{Mariani:2008zz}, MINOS \cite{Timmons:2015laq} and T2K \cite{Abe:2015awa} experiments. With the measurement of the small parameter $\sin^{2}\theta_{13}$ in the accelerator T2K \cite{Abe:2015awa}, reactor Daya Bay \cite{An:2015rpe}, RENO \cite{Kim:2014rfa} and Double Chooze \cite{Abe:2015rcp} experiments investigation of neutrino oscillations enters into a new era, era of high precision measurements. The 2015 Nobel Prize to T. Kajita  and A. McDonald  "for the discovery of neutrino oscillations, which shows that neutrinos have mass" is a very important event for the neutrino
 community which  will attract new people and give a great boost to the field.

Idea of neutrino oscillations was first proposed by B.Pontecorvo in 1957-58 soon after the theory of the two-component neutrino  was proposed \cite{Landau:1957tp} and confirmed by the Goldhaber et al experiment \cite{Goldhaber:1958nb}.  B.Pontecorvo looked in the lepton world for a phenomena analogous to $K^{0}\leftrightarrows \bar K^{0}$ oscillations. In the paper \cite{Pontecorvo:1957cp} he considered muonium ($\mu^{+}e^{-}$) to antimuonium ($\mu^{-}e^{+}$) transition. In this paper he mentioned a possibility of the neutrino oscillations. Special paper dedicated to neutrino oscillations was published by  B.Pontecorvo in 1958 \cite{Pontecorvo:1957qd}. At that time only one type of neutrino was known. B.Pontecorvo assumed that in addition to the usual weak interaction {\em exist a much weaker  interaction which does not conserve the lepton number.} Assuming maximum mixing (by the analogy with $K^{0}-\bar K^{0}$) he concluded that
``...neutrino and antineutrino are particle mixtures, i.e. symmetrical and antisymmetrical combinations of two truly neutral Majorana particles $\nu_{1}$ and $\nu_{2}$...'':
\begin{equation}\label{mix1}
|\bar\nu_{R}\rangle=\frac{1}{\sqrt{2}}(|\nu_{1}\rangle+|\nu_{2}\rangle),
\quad|\nu_{R}\rangle=\frac{1}{\sqrt{2}}(|\nu_{1}\rangle-|\nu_{2}\rangle)
\end{equation}
Here $|\bar\nu_{R}\rangle$ is the state of the right-handed antineutrino, $|\nu_{R}\rangle$ is the state of right-handed neutrino, a particle which does not take part in the weak interaction (later B.Pontecorvo proposed the name sterile for  such neutrinos), $|\nu_{1,2}\rangle$ are states of Majorana neutrinos with small masses $m_{1,2}$. As a result of the mixing (\ref{mix1}), oscillations $\bar\nu_{R}\leftrightarrows \nu_{R}$ (sterile) become possible. B.Pontecorvo discussed a possibility to check a hypothesis of neutrino oscillations in the reactor neutrino experiments. In 1958 the only known sources of neutrinos were reactors and the sun. B.Pontecorvo finished the paper \cite{Pontecorvo:1957qd} with the following remark ``...effects of transformation of neutrino into antineutrino and vice versa may be unobservable in the laboratory because of large values of $R$ (oscillation length) , but will certainly occur, at least, on an astronomic scale."

In 1962 the idea of neutrino masses and mixing was discussed by Maki, Nakagawa and Sakata \cite{Maki:1962mu}. Their proposal was based on the Nagoya model in which nucleons were considered as bound states of a vector boson and neutrino with definite mass. MNS assumed that the fields of the weak neutrinos $\nu_{e}$ and $\nu_{\mu}$ are connected with the fields of neutrinos with definite masses $\nu_{1}$ and $\nu_{2}$ (they called them true neutrinos) by the orthogonal transformation
\begin{equation}\label{mix2}
\nu_{e}=\cos\theta \nu_{1} +\sin\theta\nu_{2},\quad \nu_{\mu}=-\sin\theta \nu_{1} +\cos\theta\nu_{2}.
\end{equation}
 The phenomenon of neutrino oscillations was not considered in \cite{Maki:1962mu}. However, MNS discussed a possibility of "virtual transmutation" of $\nu_\mu$ into $\nu_e$. They estimated a time
of this transition and discussed how a possible $\nu_\mu \to \nu_e$ transition would influence the interpretation of the results of the Brookhaven experiment \cite{Danby:1962nd},\footnote{As it is well known, in this experiment it was discovered that $\nu_\mu $ and
$\nu_e$ are different particles.}
which was going on at the time when the MNS paper  was written.

In 1967 B.Pontecorvo published the second paper on neutrino oscillations \cite{Pontecorvo:1967fh}. In this paper he discussed flavor neutrino oscillations $\nu_\mu \leftrightarrows \nu_e$ and also oscillations between flavor and sterile neutrinos ($\nu_{eL} \leftrightarrows \bar\nu_{eL}$ etc). In the paper \cite{Pontecorvo:1967fh} solar neutrino oscillations were considered.  {\em Before the first results of the Davis solar neutrino experiment appeared
}, B.Pontecorvo pointed out that because of neutrino oscillations the flux of the solar $\nu_{e}$'s could be two times smaller than the expected flux. Thus, he anticipated "the solar neutrino problem".

In the Gribov and Pontecorvo paper \cite{Gribov:1968kq} it was suggested that only  active left-handed neutrinos $\nu_{e}$ and $\nu_{\mu}$ and right-handed antineutrinos $\bar\nu_{e}$ and $\bar\nu_{\mu}$ exist in nature (no sterile neutrinos). It was assumed that exist a (miliweak) interaction which does not conserve lepton numbers. After the diagonalization of such an interaction the authors came to the mixing relation
\begin{equation}\label{mix3}
\nu_{eL}=\cos\xi \phi_{1L} +\sin\xi \phi_{2L},\quad \nu_{\mu L}=-\sin\xi \phi_{1L} +\cos\xi \phi_{2L},
\end{equation}
where $\xi$ is the mixing angle and $\phi_{1}$ and $\phi_{2}$ are fields of the Majorana neutrinos with masses $m_{1}$ and $m_{1}$. They  calculated the probability of $\nu_{e}$ to survive in vacuum.
The case of the maximum mixing ($\xi=\pi/4$), analogous to the $K^{0}-\bar K^{0}$ case, was considered as the most attractive one.  Under this assumption the oscillations of solar neutrinos were discussed.

In the seventies and eighties  idea of neutrino masses and oscillations was further developed in Dubna in the papers \cite{Bilenky:1975tb}. In addition to the Gribov-Pontecorvo scheme of the neutrino mixing, based on the Majorana mass term, neutrino mixing based on the Dirac mass term and the most general  Dirac and Majorana mass term were considered. Possible reactor, accelerator, solar and atmospheric experiments on the search for neutrino oscillations were discussed. Our general point of view, which we advocated in our papers and in the first review on neutrino oscillations \cite{Bilenky:1978nj} was the following:
\begin{enumerate}
  \item There are no principles which require that neutrinos are massless particles. It is  plausible that neutrinos have small nonzero masses.
  \item Neutrino oscillations is an interference phenomenon. Search for neutrino oscillations
  is the most sensitive method to search for extremely small mass-squared differences.
  \item Experiments with neutrinos from different sources are sensitive to different neutrino mass-squared differences. Experiments on the search for neutrino oscillations must be performed with neutrinos from all existing sources.\footnote{As we know, after heroic efforts of many people this strategy led to the discovery of neutrino oscillations.}
\end{enumerate}

\section{Neutrino mixing}
Neutrino oscillations are based on the mixing of neutrino fields
\begin{equation}\label{mixing}
 \nu_{lL}(x)=\sum_{i} U_{li}\nu_{iL}(x),
\end{equation}
Here $U$ is a unitary mixing matrix and $\nu_{i}(x)$ is the field of neutrinos (Dirac or Majorana) with mass $m_{i}$.

The flavor neutrino fields $\nu_{lL}(x)$ ($l=e,\mu,\tau$) enter into the Standard Model CC and NC interactions
\begin{equation}\label{CC}
 \mathcal{L}_{I}^{CC}= -\frac{g}{\sqrt{2}}~
 j^{CC}_{\alpha}~W^{\alpha} +\rm{h.c.},\quad \mathcal{L}_{I}^{NC}= -\frac{g}{2\cos\theta_{W}}~
 j^{NC}_{\alpha}~Z^{\alpha}.
\end{equation}
Here
\begin{equation}\label{CCNC}
j^{CC}_{\alpha}=\sum_{l=e,\mu,\tau}\bar\nu_{lL} \,\gamma
_{\alpha}\, l_{L},\quad j^{NC}_{\alpha}=\sum_{l=e,\mu,\tau}\bar\nu_{lL} \,\gamma
_{\alpha}\, \nu_{lL}
\end{equation}
are charged leptonic and neutral neutrino currents.

The neutrino mixing takes place if in the total Lagrangian there is {\em a mass term nondiagonal over flavor neutrino fields}. In the case of the charged particles (leptons and quarks) only Dirac mass terms are possible. Because the electric charges of neutrinos are equal to zero three different neutrino mass terms are possible (see \cite{Bilenky:1987ty,Bilenky:2010zza}).

{\bf Dirac mass term}
\begin{equation}\label{Dmassterm}
 \mathcal{L}^{\mathrm{D}}=- \sum_{l',l=e,\mu,\tau }\bar\nu_{l' L}\,
M^{\mathrm{D}}_{l', l }\,\nu_{l R} +\rm{h.c.},
\end{equation}
where $M^{\mathrm{D}}$ is a complex, nondiagonal, $3\times 3$ matrix. After the diagonalization of the matrix $ M^{\mathrm{D}}$ we have
\begin{equation}\label{Dmixing}
 \nu_{lL}(x)=\sum^{3}_{i=1} U_{li}~\nu_{iL}(x).
\end{equation}
Here $U$ is the unitary PNMS  mixing matrix and $\nu_{i}(x)$ is the Dirac field with the mass $m_{i}$.
The Lagrangian $\mathcal{L}^{\mathrm{D}}$  conserves the total lepton number $L=L_{e}+ L_{\mu}+ L_{\tau}$. Neutrino $\nu_{i}$ and antineutrino $\bar\nu_{i}$ differ by the lepton number: $L(\nu_{i})=1,~~L(\bar\nu_{i})=-1$.

{\bf Majorana mass term}
\begin{equation}\label{Mjmassterm}
\mathcal{L}^{\mathrm{M}}=-\frac{1}{2}\, \sum_{l', l=e,\mu,\tau
}\bar\nu_{l' L}\, M^{\mathrm{M}}_{l' l  }\,(\nu_{l L})^{c}
+\rm{h.c.},
\end{equation}
where $M^{\mathrm{M}}$ is a complex, nondiagonal, symmetrical $3\times 3$ matrix and $(\nu_{l L})^{c}=C\bar\nu^{T}_{l L}$ is the conjugated field. The mass term (\ref{Mjmassterm}) violates not only flavor lepton numbers  but also the total lepton number $L$. After the diagonalization of the matrix  $M^{\mathrm{M}}$ we have
\begin{equation}\label{Mjmixing}
 \nu_{lL}(x)=\sum^{3}_{i=1} U_{li}~\nu_{iL}(x).
\end{equation}
Here $U$ is a unitary $3\times 3$ mixing matrix and
\begin{equation}\label{Mj}
\nu_{i}(x)=\nu^{c}_{i}(x)
\end{equation}
is the Majorana field with the mass  $m_{i}$  ($\nu_{i}\equiv \bar\nu_{i})$.

The most general {\bf Dirac and Majorana mass term}
\begin{equation}\label{DMjmassterm}
\mathcal{L}^{\mathrm{D+M}}=\mathcal{L}^{\mathrm{M}}+\mathcal{L}^{\mathrm{D}}-\frac{1}{2}\, \sum_{s', s=s_{1},...s_{n_{s}}
}\overline{(\nu_{s' R})^{c}}\, M^{\mathrm{R}}_{s' s  }\,\nu_{s R}
+\rm{h.c.}
\end{equation}
($M^{\mathrm{R}}$ is a complex symmetrical matrix) violates lepton numbers and require left-handed and right-handed neutrino fields. After the diagonalization of the mass term $\mathcal{L}^{\mathrm{D+M}}$ we find
\begin{equation}\label{DMjmixing}
 \nu_{lL}(x)=\sum^{3+n_{s}}_{i=1} U_{li}~\nu_{iL}(x),\quad    (\nu_{sR}(x))^{c}=\sum^{3+n_{s}}_{i=1} U_{s i}~\nu_{iL}(x).
\end{equation}
Here $U$ is a unitary $(3+n_{s})\times(3+n_{s})$ matrix and $\nu_{i}(x)=\nu^{c}_{i}(x)$ is the field of a Majorana lepton with definite mass.

The mixing (\ref{DMjmixing}) open different possibilities: the seesaw possibility of the generation of small neutrino masses \cite{seesaw}, a possibility of transitions of flavor neutrinos into sterile states etc.

Let us notice that the Dirac mass term can be generated by the standard Higgs mechanism. The Majorana and the Dirac and Majorana mass terms can be generated only by a beyond the SM mechanisms.

\section{Flavor neutrino states}

There exist different methods of the derivation (of the same) expression for transition probabilities. We will present here a method based on the notion of {\em the coherent flavor neutrino states} (see \cite{Bilenky:2010zza})
\begin{equation}\label{flavstate}
|\nu_{l}\rangle =\sum_{i}U^{*}_{li}~|\nu_{i}\rangle,\quad l=e,\mu,\tau
\end{equation}
Here $|\nu_{i}\rangle$ is the state of neutrino (Dirac or Majorana) with mass $m_{i}$, momentum $\vec{p}$ and energy
$E_{i}=\sqrt{p^{2}+m^{2}_{i}}\simeq E+\frac{m^{2}_{i}}{2E}~(E=p)$, and
$|\nu_{l}\rangle$  is the state the flavor neutrino $\nu_{l}$ which is produced together with $l^{+}$ in a CC weak decay ($\pi^{+}\to \mu^{+}+\nu_{\mu}$ etc) or produces   $l^{-}$ in a CC neutrino reaction ($\nu_{\mu}+N\to \mu^{-}+X$ etc).

The relation (\ref{flavstate}) is  valid if neutrino mass-squared differences are so small that  in weak decays production of neutrinos with different masses can not be resolved. It follows from the Heisenberg uncertainty relation that this condition is satisfied in neutrino oscillation experiments with neutrino energies many orders of magnitude larger than neutrino masses.

The possibility to resolve small neutrino mass-squared differences is based on the time-energy uncertainty relation (see \cite{Bilenky:2011pk})
\begin{equation}\label{timeenergy1}
 \Delta E~ \Delta t \gtrsim 1.
\end{equation}
Here $\Delta t$ is a time interval during which the state
with the energy uncertainty  $\Delta E$ is significantly changed. In the case of neutrino beams from (\ref{timeenergy1}) we find
\begin{equation}\label{timeenergy4}
 |\Delta m^{2}_{ki}|~\frac{L}{2E}\gtrsim 1,
\end{equation}
where $L\simeq \Delta t$ is the distance between a neutrino source and neutrino detector. For "atmospheric" and "solar" mass-squared differences $\Delta m^{2}_{A}\simeq 2.4\cdot 10^{-3}~\mathrm{eV}^{2}$ and
$\Delta m^{2}_{S}\simeq 7.5\cdot 10^{-5}~\mathrm{eV}^{2}$ the condition (\ref{timeenergy4}) is satisfied in the atmospheric Super-Kamiokande  \cite{Fukuda:1998ah} , long baseline accelerator K2K \cite{Mariani:2008zz}, MINOS  \cite{Timmons:2015laq}, T2K \cite{Abe:2015awa}, reactor
KamLAND \cite{Eguchi:2002dm}, Daya Bay \cite{An:2015rpe}, RENO \cite{Kim:2014rfa} Double Chooze \cite{Abe:2015rcp} and other neutrino oscillation experiments.

We will finish this section with a remark about the states of sterile neutrinos which (by definition) do not
interact with leptons and quarks via the SM interaction. If in addition to the flavor neutrinos $\nu_{l}$
 sterile neutrinos $\nu_{s}$ exist, their states  {\em are determined} as follows
\begin{equation}\label{sterilestate}
|\nu_{s}\rangle =\sum^{3+n_{s}}_{i=1}U^{*}_{si}~|\nu_{i}\rangle,\quad s=s_{1},s_{1},...
\end{equation}
where $U$ is a unitary $(3+n_{s})\times (3+n_{s})$ matrix. The states of active and sterile neutrinos (\ref{flavstate}) and (\ref{sterilestate}) satisfy the condition
\begin{equation}\label{normaliz}
\langle \alpha'|\alpha\rangle=\delta_{\alpha'\alpha},\quad \alpha',\alpha=e,\mu,\tau,s_{1},s_{1},...s_{n_{s}}.
\end{equation}
Neutrino oscillations is a direct consequence of the fact that flavor (and sterile) neutrinos are described by coherent states (\ref{flavstate}) and (\ref{sterilestate}).

\section{Neutrino oscillations in vacuum}
Let us assume that at the initial time $t=0$ a flavor neutrino $\nu_{\alpha}$ was produced. In the general case of flavor and sterile neutrinos at the time $t$ we have
\begin{equation}\label{evolution}
|\nu_{\alpha}\rangle_{t}=e^{-iH_{0}t}~|\nu_{\alpha}\rangle=\sum^{3+n_{s}}_{i=1}|\nu_{i}\rangle~e^{-iE_{i}t}~ U^{*}_{\alpha i}=\sum_{\alpha'}|\alpha'\rangle(\sum^{3+n_{s}}_{i=1}U_{\alpha'i}~e^{-iE_{i}t}~ U^{*}_{\alpha i}).
\end{equation}
Thus, for the  $\nu_{\alpha}\to \nu_{\alpha'}$ transition probability we find
\begin{equation}\label{Transition}
P(\nu_{\alpha}\to \nu_{\alpha'})=|\sum^{3+n_{s}}_{i=1}U_{\alpha'i}~e^{-iE_{i}t}~ U^{*}_{\alpha i}|^{2}
\end{equation}
We will present here  convenient expression for $\nua{\alpha}\to
\nua{\alpha'}$ transition probability (see \cite{Bilenky:2015xwa}). From (\ref{Transition}) we have
\begin{equation}\label{transition}
P(\nu_{\alpha}\to \nu_{\alpha'})=|\sum^{3+n_{s}}_{i=1}U_{\alpha'i}~e^{-2i\Delta_{pi}}~ U^{*}_{\alpha i}|^{2}=
|\delta_{\alpha'\alpha}-2i\sum_{i}U_{\alpha'i}~e^{-i\Delta_{pi}}\sin\Delta_{pi}~U^{*}_{\alpha i}|^{2}
\end{equation}
where $p$ is arbitrary, fixed index and
\begin{equation}\label{standard1}
\Delta_{pi}=\frac{\Delta m^{2}_{pi}L}{4E}, \quad \Delta m^{2}_{pi}=m^{2}_{i}-m^{2}_{p}.
\end{equation}
Let us notice that in Eq. (\ref{transition})
\begin{itemize}
  \item $i\neq p$,
  \item we extract the common phase $e^{\frac{-im^{2}_{p}L}{2E}}$,
  \item we used the unitarity condition $\sum _{i}U_{\alpha'i}~U^{*}_{\alpha i}=\delta_{\alpha'\alpha}$.
\end{itemize}
From (\ref{transition}) we find
\begin{eqnarray}\label{transition2}
&&P(\nua{\alpha}\to
\nua{\alpha'})=\delta_{\alpha'\alpha}-4~\sum_{i}|U_{\alpha
i}|^{2}(\delta_{\alpha'\alpha}-|U_{\alpha'i}|^{2})~\sin^{2}\Delta_{pi}
\nonumber\\
&&+8~\sum_{i>k}[\mathrm{Re}~(U_{\alpha' i}U^{*}_{\alpha i}U_{\alpha'
k}^{*}U_{\alpha k})~\cos(\Delta_{pi}-\Delta _{pk}) \nonumber\\
&&\pm ~\mathrm{Im}~(U_{\alpha' i}U^{*}_{\alpha i}U_{\alpha'
k}^{*}U_{\alpha k})~\sin(\Delta_{pi}-\Delta _{pk})] \sin\Delta
_{pi}\sin\Delta_{pk}.
\end{eqnarray}
Here  +(-) sign refer to $\nu_{\alpha}\to
\nu_{\alpha'}$ ($\bar\nu_{\alpha}\to \bar\nu_{\alpha'}$) transition.

From our point of view there are some advantages of the expression (\ref{transition2}) with respect to the standard expression (for the standard expression see \cite{Giunti:2007ry}).
\begin{enumerate}
  \item Only independent mass-squared differences enter into this expression.
  \item The unitarity condition is fully implemented in (\ref{transition2}). As a result  only independent terms enter into this expression.
\end{enumerate}
We will consider now the most important case of the three-neutrino mixing. Usually neutrino masses are labeled in such a way that $m_{2}>m_{1}$ and solar ("small") mass-squared difference is determined as follows
\begin{equation}\label{Smass}
 m_{2}^{2}- m_{1}^{2} =\Delta m_{12}^{2} \equiv\Delta m_{S}^{2}.
\end{equation}
For the neutrino mass spectrum there are two possibilities:
\begin{enumerate}
  \item Normal spectrum (NS) : $\Delta m_{S}^{2}$ is the  difference between square of masses of the lightest neutrinos. In this case $m_{3}>m_{2}>m_{1}$.
  \item Inverted spectrum (IS): $\Delta m_{S}^{2}$ is the  difference between square of masses of the heaviest neutrinos. In this case $m_{2}>m_{1}>m_{3}$.
  \end{enumerate}
We will determine the atmospheric ("large") neutrino mass squared difference in the following way
\begin{equation}\label{Amass}
NS: \Delta m_{A}^{2}=\Delta m_{23}^{2},\quad IS:  \Delta m_{A}^{2}=|\Delta m_{13}^{2}|.
\end{equation}
Let us notice that there exist different  definition of this quantity in the literature
\begin{enumerate}
  \item The Bari group \cite{Capozzi:2015uma} determines atmospheric mass-squared difference as follows
  \begin{equation}\label{Bari}
(\Delta m_{A}^{2})'= \frac{1}{2}|\Delta m_{13}^{2}+\Delta m_{23}^{2}|=\Delta m_{A}^{2}+\frac{1}{2}\Delta m_{S}^{2}.
  \end{equation}
  \item The NuFit group \cite{Gonzalez-Garcia:2014bfa} determines atmospheric mass-squared difference in the following way
 \begin{equation}\label{NuFit}
 (\Delta m_{A}^{2})''=\Delta m_{13}^{2}~(NS)= |\Delta m_{23}^{2}|~ (IS)=\Delta m_{A}^{2}+\Delta m_{S}^{2}.
    \end{equation}
\item In \cite{Nunokawa:2005nx} the parameter $\Delta m^{2}_{ee}$ was introduced. It is determined as follows
\begin{equation}\label{Daya}
\Delta m^{2}_{ee}=\cos^{2}\theta_{12}\Delta m^{2}_{13}+\sin^{2}\theta_{12}\Delta m^{2}_{23}
\end{equation}
The parameter $\Delta m^{2}_{ee}$ is connected with $\Delta m_{A}^{2}$ and $\Delta m_{S}^{2}$ by the relations
\begin{equation}\label{Daya1}
 \Delta m^{2}_{ee}=\Delta m^{2}_{A}+\cos^{2}\theta_{12}\Delta m^{2}_{S}~(NS),~
 | \Delta m^{2}_{ee}|=\Delta m^{2}_{A}+\sin^{2}\theta_{12}\Delta m^{2}_{S}~(IS).
 \end{equation}
\end{enumerate}
As it is seen from (\ref{Amass}),  (\ref{Bari}) (\ref{NuFit}) (\ref{Daya1}) different definitions of "large" mass-squared difference differ only by a few \%. However, neutrino oscillation experiments  enter now into
precision era when neutrino oscillation parameters will be measured with \% accuracy. We believe that the consensus
in definition of "large" neutrino mass-squared difference must be found.

For the probability of the transition $\nua{l}\to \nua{l'}$ ($l,l'=e,\mu,\tau$)
in the case of normal and inverted mass spectra from (\ref{transition2}) we find, correspondingly, the following expressions

\begin{eqnarray}
&&P^{NS}(\nua{l}\to \nua{l'})
=\delta_{l' l }
-4|U_{l 3}|^{2}(\delta_{l' l} - |U_{l' 3}|^{2})\sin^{2}\Delta_{A}\nonumber\\&&-4|U_{l 1}|^{2}(\delta_{l' l} - |U_{l' 1}|^{2})\sin^{2}\Delta_{S}
-8~[\mathrm{Re}~(U_{l' 3}U^{*}_{l 3}U^{*}_{l'
1}U_{l 1})\cos(\Delta_{A}+\Delta_{S})\nonumber\\
&&\pm ~\mathrm{Im}~(U_{l' 3}U^{*}_{l 3}U^{*}_{l'
1}U_{l 1})\sin(\Delta_{A}+\Delta_{S})]\sin\Delta_{A}\sin\Delta_{S},
\label{Genexp5}
\end{eqnarray}
and
\begin{eqnarray}
&&P^{IS}(\nua{l}\to \nua{l'})
=\delta_{l' l }
-4|U_{l 3}|^{2}(\delta_{l' l } - |U_{l' 3}|^{2})\sin^{2}\Delta_{A}\nonumber\\&&-4|U_{l 2}|^{2}(\delta_{l' l} - |U_{l' 2}|^{2})\sin^{2}\Delta_{S}
-8~[\mathrm{Re}~(U_{l' 3}U^{*}_{l 3}U^{*}_{l'
2}U_{l 2})\cos(\Delta_{A}+\Delta_{S})\nonumber\\
&&\mp ~\mathrm{Im}~(U_{l' 3}U^{*}_{l 3}U^{*}_{l'
2}U_{l 2})\sin(\Delta_{A}+\Delta_{S})]\sin\Delta_{A}\sin\Delta_{S}.
\label{Genexp6}
\end{eqnarray}
The transition probabilities (\ref{Genexp5}) and (\ref{Genexp6}) are  the sum of  atmospheric,  solar  and interference terms. Notice that expression (\ref{Genexp6}) can be obtained from (\ref{Genexp5})
by the change $U_{l 1}\to U_{l 2}$ and  $(\pm)\to (\mp)$ in the last term.

 The values of the oscillation parameters obtained from global analysis of existing data by the NuFit group \cite{Gonzalez-Garcia:2014bfa} are presented in the Table 1.
\begin {table}[H]
\caption {Values of neutrino oscillation parameters obtained in \cite{Gonzalez-Garcia:2014bfa} from the global fit of existing data} \label{tab:title}
\begin{center}
\begin{tabular}{|c|c|c|}
  \hline  Parameter &  Normal Spectrum& Inverted Spectrum\\
\hline   $\sin^{2}\theta_{12}$& $0.304^{+0.013}_{-0.012}$& $0.304^{+0.013}_{-0.012}$
\\
\hline    $\sin^{2}\theta_{23}$& $0.452^{+0.052}_{-0.028}$& $ 0.579^{+0.025}_{-0.037}$
\\
\hline   $\sin^{2}\theta_{13}$ & $ 0.0218^{+0.0010}_{-0.0010}$&  $0.0219^{+0.0011}_{-0.0010}$
\\
\hline   $\delta $~(in $^{\circ}$) & $(306^{+39}_{-70})$& $ (254^{+63}_{-62})$
\\
\hline $\Delta m^{2}_{S}$& $(7.50^{+0.19}_{-0.17})\cdot 10^{-5}~\mathrm{eV}^{2}$&$(7.50^{+0.19}_{-0.17})\cdot 10^{-5}~\mathrm{eV}^{2}$\\
\hline $\Delta m^{2}_{A}$& $(2.457^{+0.047}_{-0.047})\cdot 10^{-3}~\mathrm{eV}^{2}$&$(2.449^{+0.048}_{-0.047})\cdot 10^{-3}~\mathrm{eV}^{2}$\\
\hline
\end{tabular}
\end{center}
\end{table}
\section{Neutrino and the Standard Model}
After the discovery of the Higgs boson at LHC  the Standard Model acquired  the status  of the theory of elementary particles in the electroweak range (up to $\sim$ 300 GeV). The Standard Model is based on the following principles:
\begin{itemize}
  \item Local gauge symmetry.
  \item Unification of the  weak and electromagnetic interactions.
  \item Spontaneous breaking of the electroweak symmetry.
\end{itemize}
It was suggested in \cite{Bilenky:2014ema} that in the framework of these principles nature choose the simplest, most economical possibilities. The Standard Model started with the theory of the two-component neutrino. The two-component, massless, Weil neutrino is the {\em simplest possibility for the particle with spin 1/2}:
only two degrees of freedom. The local $SU_{L}(2)$ group with the lepton doublets
\begin{eqnarray}\label{doublets}
\psi^{lep}_{eL}=\left(
\begin{array}{c}
\nu'_{eL} \\
e'_L \\
\end{array}
\right),~\psi^{lep}_{\mu L}=\left(
\begin{array}{c}
\nu'_{\mu L} \\
\mu'_L \\
\end{array}
\right),~\psi^{lep}_{\tau
 L}=\left(
\begin{array}{c}
\nu'_{\tau L} \\
\tau'_L \\
\end{array}
\right)
\end{eqnarray}
and corresponding quark doublets is {\em the simplest possibility} which allows to include  charged leptons and quarks in addition to neutrinos.

In order to unify weak and electromagnetic interactions we need to enlarge the symmetry group: in electromagnetic currents of {\em charged particles}  enter left-handed and right-handed fields. The simplest enlargement is  the $SU_{L}(2)\times U_{Y}(1)$ group where  $U_{Y}(1)$ is the group of the weak hypercharge $Y$
 determined by the Gell-Mann-Nishijima relation $Q=T_{3}+\frac{1}{2}Y$. Neutrinos have no electromagnetic interaction. {\em Unification of the weak and electromagnetic interactions
does not require right-handed neutrino fields.} The SM interaction of leptons, neutrinos and quarks with gauge vector bosons {\em is the minimal interaction} compatible with the local $SU_{L}(2)\times U_{Y}(1)$ invariance.

The SM mechanism  of the mass generation is the Brout-Englert-Higgs mechanism based on the assumption of the existence of scalar Higgs fields. In order to generate masses of  $W^{\pm}$ and $Z^{0}$ bosons we need to have three (Goldstone) degrees of freedom. {\em Minimal possibility is a doublet of complex Higgs fields} (four degrees of freedom). With this assumption one scalar, neutral Higgs boson is predicted. This prediction is in a good agreement with existing LHC data.

Masses of  $W^{\pm}$ and $Z^{0}$ bosons are given in the SM by the relations
\begin{equation}\label{WZmass}
m_{W}=\frac{1}{2}g~v,\quad m_{Z}=\frac{1}{2}\sqrt{g^{2}+g'^{2}}~v=\frac{g}{2\cos\theta_{W}}v,
\end{equation}
where $v=(\sqrt{2}G_{F})^{-1/2}=246~~ \mathrm{GeV}$ is the parameter which characterizes the scale of the electroweak symmetry breaking. Lepton and quark masses and mixing are due to  $SU_{L}(2)\times U_{Y}(1)$ invariant Yukawa interactions which generate Dirac mass terms. For the charged leptons we have
\begin{equation}\label{Dirmass}
\mathcal{L}^{lep}_{Y}=-\sum_{l}m_{l}\bar l~ l,
\end{equation}
where $m_{l}=y_{l}~v$ and $y_{l}$ is the Yukawa constant. {\em Neutrinos in the minimal SM after spontaneous breaking of the electroweak symmetry remain two-component, massless, Weyl particles.}

\section{The Weinberg mechanism of the neutrino mass generation}
In the framework of the minimal SM  neutrino masses and mixing can be  generated only by a beyond the SM mechanism. The most general method which allows to describe effects of a beyond the SM physics is the method of the effective Lagrangian. The effective Lagrangian is a $SU_{L}(2)\times U_{Y}(1)$ invariant,
dimension five or more local operator built from SM fields. In order to built the effective Lagrangian which generate a neutrino mass term we must use the lepton doublets
(\ref{doublets}) and the Higgs doublet
\begin{eqnarray}\label{Higgs}
\phi=\left(
\begin{array}{c}
\phi_{+} \\
\phi_{0} \\
\end{array}
\right)
\end{eqnarray}
The effective Lagrangian which generate the neutrino mass term has the form \cite{Weinberg:1979sa}
\begin{equation}\label{Weinberg}
 \mathcal{L}_{I}^{\mathrm{eff}}=-\frac{1}{\Lambda}~\sum_{l_{1},l_{2}}(\bar \psi^{lep}_{l_{1}L}\tilde{\phi })~ Y_{l_{1}l_{2}}~(\tilde{\phi }^{T} (\psi^{lep}_{l_{2}L})^{c})+\mathrm{h.c.},
\end{equation}
where the parameter $\Lambda$  characterizes a scale of a beyond the SM physics ($\Lambda\gg v$) and $\tilde{\phi}=i\tau_{2}\phi^{*}$ is the conjugated doublet. Let us stress that the Lagrangian (\ref{Weinberg}) does not conserve the total lepton number.\footnote{The Lagrangian (\ref{Weinberg}) can be generated (in the second order of the perturbation theory) by the seesaw interaction of the Higgs-lepton pair with a heavy Majorana right-handed lepton.}

After spontaneous symmetry breaking from (\ref{Weinberg}) we come to the Majorana mass term
\begin{equation}\label{Majgenerated}
\mathcal{L}^{\mathrm{M}}= -\frac{1}{2}\,\frac{v^{2}}{\Lambda}\sum_{l_{1},l_{2}}
\bar\nu'_{l_{1}L}\,Y_{l_{1}l_{2}} ( \nu'_{l_{2}L})^{c}+\mathrm{h.c.}=-\frac{1}{2}\sum^{3}_{i=1}m_{i}~\bar \nu_{i}\nu_{i}.
\end{equation}
Here $\nu_{i}= \nu^{c}_{i}$ is the field of the neutrino Majorana with the mass
\begin{equation}\label{Majgenerated2}
 m_{i}=\frac{v^{2}}{\Lambda}y_{i}=\frac{v}{\Lambda}(y_{i}v),
\end{equation}
where $y_{i}$ is a Yukawa coupling. In (\ref{Majgenerated2}) $y_{i}v$ is a "typical" fermion mass in SM. Thus, neutrino masses, generated by the effective Lagrangian (\ref{Weinberg}), are suppressed with respect to "SM masses" by a factor
$$\frac{v}{\Lambda}=\frac{\mathrm{scale~ of~ SM}}{\mathrm{scale ~of ~a ~new~ physics}}\ll 1$$

The mechanism  we have considered is, apparently,{\em the most economical and natural beyond the SM mechanism of the neutrino mass generation.} There are two general consequences of this mechanism.
\begin{itemize}
  \item Neutrinos with definite masses $\nu_{i}$ are Majorana particles.
  \item The number of neutrinos with definite masses is equal to the number of lepton-quark generations (three). This means that in this scheme there are no transitions of flavor neutrinos into sterile states.
 \end{itemize}
The study of the lepton number violating neutrinoless double $\beta$-decay ($0\nu\beta\beta$-decay)
\begin{equation}\label{bb}
(A,Z)\to (A,Z+2) +e^{-}+e^{-}
\end{equation}
of some even-even nuclei is most sensitive way to investigate the Majorana nature of neutrinos with definite masses (see review \cite{Bilenky:2014uka}). The probability of the process (\ref{bb}) is proportional to square of the Majorana neutrino mass
\begin{equation}\label{bb1}
m_{\beta\beta}=\sum_{i}U^{2}_{ei}m_{i}
\end{equation}
and is very small. It has the following general form
\begin{equation}\label{bb2}
\frac{1}{T^{0\nu}_{1/2}}=|m_{\beta\beta}|^{2}~|M^{0\nu}|^{2}~
G^{0\nu}(Q,Z).
\end{equation}
Here $M^{0\nu}$ is the nuclear matrix element and $G^{0\nu}(Q,Z)$
is known phase factor.

Several experiments on the search for the $0\nu\beta\beta$  of different nuclei are going on and are in preparation. Up to now the process was not observed. From the data of recent experiments EXO-200 \cite{Albert:2014awa}, KamLAND-Zen \cite{Asakura:2014lma} and GERDA \cite{Wester:2015uiu} the following upper bounds were, correspondingly, obtained
\begin{equation}\label{bb3}
|m_{\beta\beta}|<(1.9-4.5)\cdot 10^{-1}~\mathrm{eV},~~(1.4-2.8)\cdot 10^{-1}~\mathrm{eV},~~(2-4)\cdot 10^{-1}~\mathrm{eV}
\end{equation}
In future experiments on the search for $0\nu\beta\beta$ decay the values   $|m_{\beta\beta}|\simeq \mathrm{a~few}\cdot 10^{-2}~\mathrm{eV}$ are planned to be reached.

Indications in favor of transitions of flavor neutrinos into sterile states were obtained in the LSND \cite{Aguilar:2001ty} and MiniBooNE
short baseline accelerator experiments and in the   GALLEX and SAGE calibration experiments and in short baseline reactor experiments which were  reanalyzed with a
new  reactor antineutrino flux (see recent review \cite{Gariazzo:2015rra}). Many new short baseline source, reactor and accelerator neutrino experiments on the search for sterile neutrinos with masses $\sim 1$ eV are in preparation (see \cite{Lasserre:2014ita}). There is no doubt that in a few years the sterile neutrino anomaly will be resolved.

\section{Conclusion}
Neutrino masses and mixing, discovered via the observation of neutrino oscillations, is a first particle physics evidence of a new beyond the SM physics. We discuss here briefly first proposals for neutrino oscillations and first steps in the development of the theory of neutrino oscillations. Then we consider basics of neutrino mixing and oscillations, convenient formalism  for neutrino oscillations in vacuum and  the definition of the atmospheric neutrino mass-squared difference. In the final part of the paper we discuss the important role which play two-component neutrino in the Standard Model and the most economical Weinberg mechanism of the generation of the Majorana neutrino masses.

I acknowledge the support of RFFI grant 16-02-01104.

\end{document}